\documentclass[10pt,letterpaper]{article}
\usepackage[top=0.85in,left=2.75in,footskip=0.75in]{geometry}

\usepackage{amsmath,amssymb}

\usepackage{changepage}

\usepackage[utf8x]{inputenc}

\usepackage{textcomp,marvosym}

\usepackage{cite}

\usepackage{nameref,hyperref}

\usepackage{microtype}
\DisableLigatures[f]{encoding = *, family = * }

\usepackage[table]{xcolor}

\usepackage{array}

\newcolumntype{+}{!{\vrule width 2pt}}

\newlength\savedwidth

\raggedright
\usepackage[aboveskip=1pt,labelfont=bf,labelsep=period,justification=raggedright,singlelinecheck=off]{caption}

\makeatletter
\renewcommand{\@biblabel}[1]{\quad#1.}
\makeatother

\usepackage{lastpage,fancyhdr,graphicx}
\usepackage{epstopdf}
\fancyheadoffset[L]{2.25in}
\fancyfootoffset[L]{2.25in}
\begin{document}
\vspace*{0.2in}

\begin{flushleft}
{\Large
\textbf\newline{Registered Report : Perception of Other's Musical Preferences Based on Their Personal Values} % Please use "sentence case" for title and headings (capitalize only the first word in a title (or heading), the first word in a subtitle (or subheading), and any proper nouns).
}
\newline
\\
Sandy Manolios*,
Catholijn M. Jonker,
Cynthia C.S. Liem
\\
\bigskip
Department of Intelligent Systems, Delft University of Technology, Netherlands
\\
\bigskip

% Insert additional author notes using the symbols described below. Insert symbol callouts after author names as necessary.
% 
% Remove or comment out the author notes below if they aren't used.
%
% Primary Equal Contribution Note
%\Yinyang These authors contributed equally to this work.

% Additional Equal Contribution Note
% Also use this double-dagger symbol for special authorship notes, such as senior authorship.
%\ddag These authors also contributed equally to this work.

%\textcurrency Current Address: Dept/Program/Center, Institution Name, City, State, Country % change symbol to "\textcurrency a" if more than one current address note
% \textcurrency b Insert second current address 
% \textcurrency c Insert third current address


% Group/Consortium Author Note
%\textpilcrow Membership list can be found in the Acknowledgments section.

% Use the asterisk to denote corresponding authorship and provide email address in note below.
* s.manolios@tudelft.nl

\end{flushleft}
\section*{Abstract}

%Music is as present in our society as it is diverse. Different individuals enjoy different types of music. %But why is that? 
The present study is part of a research line seeking to uncover the mysteries of what lies behind people's musical preferences in order to provide better music recommendations. More specifically, it takes the angle of personal values. Personal values are what we as people strive for, and are a popular tool in marketing research to understand customer preferences for certain types of product. Therefore, it makes sense to explore their usefulness in the music domain.
Based on a previous qualitative work using the Means-End theory, we designed a survey in an attempt to more quantitatively approach the relationship between personal values and musical preferences. We support our approach with a simulation study as a tool to improve the experimental procedure and decisions.

%\linenumbers

% Use "Eq" instead of "Equation" for equation citations.
\section{Introduction}

%Everyone enjoys music, but we don't enjoy the same ones. Out there, there is a myriad of very different types of music. You probably love some of them, hate some others, and are rather indifferent to the rest of them. But there are also a lot you never heard before. So how could you know whether you will like them or not? Recommender systems are made just for this, helping you to find new music you will enjoy. But they often focus on what you previously listened to, mostly directing you towards similar songs. Some works have focused on improving recommender systems by directing them towards more surprising, but still good, recommendations. This paper belongs to this research line and aims to understand better the links between musical preferences and psychological characteristics in order to provide better and taste expanding recommendations. 

Everyone enjoys music, but a myriad of very different types of music exist out there. Not everyone enjoys the same ones, and trying to find the reasons behind people's musical preferences is not a new question in science. 
Two research fields investigated it: psychology and marketing. Psychology has looked at the relationship between musical preferences and psychological characteristics or lifestyles \cite{north2007lifestyle1,north2007lifestyle2,north2007lifestyle3,laplante2014improving,laplante2012influence,dunn2012toward,schafer2017can}. But to the best of our knowledge, it has not considered personal values yet, with one only exception \cite{manolios2019influence}. Personal values are a psychological characteristic that represents what is most important for people in life. The marketing field has long realized their potential to predict which product attributes consumers would be attracted to. 
Understanding what makes people like the music they like has become increasingly interesting with the development of new technologies. The streaming era allows almost anyone to access all those different kind of songs without going through manually curated collections such as physical music stores and radio stations. A better understanding of musical taste could, for example, help to improve technologies designed to help users navigate through those ever-increasing music collections. 
%
%In a previous work \cite{manolios2019influence}, we 
\cite{manolios2019influence} used a traditional marketing technique to connect a group of participant's musical preferences and values through interviews. They gathered some qualitative insights and built a map. But, while informative, those qualitative insights cannot be generalized. To pursue on this line of work, we will conduct a survey study based on this previous work to better understand the interplay between personal values and musical preferences by taking a slightly different angle with a larger pool of participants.

\subsection{Musical Preferences and Psychological Characteristics} \label{musicpsycho}

The field of psychology has conducted extensive research in an attempt to tie musical preferences to a wide range of psychological characteristics. \cite{laplante2014improving} summarizes most of their findings. It includes political orientation, religious beliefs, gender, ethnicity, age, education and the social influences of peers and parents. Musical preferences are indeed strongly influenced by what musics people are exposed to during their teenage years \cite{laplante2012influence}.  
Musical preferences were also investigated in the prism of lifestyle through an extensive survey. \cite{north2007lifestyle1,north2007lifestyle2,north2007lifestyle3}

One of the most widely psychological characteristic that have been studied in connection with musical preferences is personality \cite{dunn2012toward}. A more recent meta-analysis \cite{schafer2017can} however concluded that the effects were quite small and that therefore "personality traits barely account for interindividual differences in music preferences". Though personality has been extensively studied, there are very little works, to the best of our knowledge, that considered personal values in their potential relationship with music preferences.

%There are some limitations to those studies that are worth mentioning. Most of them rely on participants' declared musical preferences. \cite{dunntowards} showed that this is slightly different from what people actually consume, though both are strongly correlated. Musical preferences are often gathered by genre (such as pop, rock, blues, metal,...). But it is now widely acknowledge in the MIR (Music Information Retrieval) field that genre is very subjective, even if it is the most common way to classify music, as well as the most understandable one for the general public \cite{renfrowGosling}.

\subsection{Personal Values}

Personal values are a core component of identity, representing who people want to be and pointing to what is most important to them in life. People will be intrinsically motivated to consume products that promote their own personal values, and even \emph{not} to consume products that demote them. Psychology has been studying personal values for decades. As they are a subjective constructs, many different models have been built. One of the most famous one is the Schwartz's model \cite{schwartz2006basic}, which itself had been through several iterations. In this paper, we used the version called Portrait Value Survey (PVQ) \cite{Schwartz2001,Schwartz2005}. It is composed of 11 values falling in 4 categories: Conservation (caring about one's safety in every aspect of one's life), Openness to Change (caring about independence and discovery), Self-Transcendence (caring for the world) and Self-Enhancement (caring for oneself). The values and the categories are organized by motivational similarities.

%\section{Previous Works}
\subsection{Means-End Theory}

Psychology is not the only field to have taken an interest in personal values. The marketing field found that many consumer choices are motivated by their personal values: e.g., whether to pick the cheapest option, the famous brand or the environmental-friendly product. Even the size or material of a wine cooler bottle can be a factor of value-based preference \cite{reynolds1988laddering}. This is based on the Means-End theory \cite{gutman1982means} according to which characteristics of products are a means for consumers to achieve an end: their personal values.

%The Means-Ends theory suggest that personal value-aware recommendation would lead to useful personalization strategies in the RS domain. However, the role of personal values in RS has been understudied so far, although the few existing works indeed support the idea that personal values have concrete potential to improve recommendations \cite{srivastava2017transfer,liu2019personality}. To study this potential in more depth, it will be necessary to make associations between personal values and consumer preferences at scale. 
So far, to the best of our knowledge, these associations have mostly been made in a more manual and qualitative way, through a marketing interviewing technique called the laddering  technique~\cite{reynolds1988laddering,saaka2004laddering}. During the interviews, the interviewer started by getting from the participant a certain number of concrete characteristics of the product being evaluated, such as being sold in a glass bottle for soda or containing less sugar. Those elements are called Attributes. From there, the interviewer asks "What makes that important to you?" or "Why is it important to you?" questions to go up levels of abstraction up to the interviewee's Values. The middle steps between Attribute and Values are called Consequences. It can be regarded as the effect the consumer is trying to get from the product attribute, such as looking sophisticated or be more productive. 

A literature review from 1994 \cite{van1994domain} claimed that this Consequences' middle step is necessary, because at a larger scale, it strongly connects to both Attributes and Values, which are only weakly connected to each other.\footnote{Unfortunately, we could not find the study on which this claim was based.}

This technique was e.g.\ used to elicit associations between preferences and values for smartphones \cite{leitner2008mobile}, housing \cite{collen2001values} and ethical clothing \cite{jagel2012individual}.

\subsection{Means-End Theory and Musical Preferences}

Another of those interviews studies focused on music \cite{manolios2019influence}, sharing some valuable insights about the relationship between personal values and music. 
%Because of the format, those relationships were elicited in the users' perspective about their own behaviors. In other words, their self-representations of how their own musical preferences are linked to their values. 
This work aims to extend this study on two major points. The first one is to use a different angle to look at those values and musical preferences relationships by asking participants about other people's musical preferences instead of their own. This approach will complement the angle of the first study : how people perceive the relationship between their own values and preferences.

The second one is to use a more quantitative method %that will allow generalizing the knowledge about those links brought by the study
with a focus on data simulation as a mean to reinforce the user experiment. This will make this exploratory study a blend of a qualitative and quantitative approach. the results will then be more, though not entirely, generalizable. 
%By doing so, this work will also contribute by proposing a more quantitative way to use the Means-End Theory.

\subsection{Research Questions}

This work seeks to look more quantitatively at the findings from a previous qualitative study by investigating two main research questions.

\begin{itemize}

\item Research Question 1: What are the links between values, consequences and musical attributes? (Exploratory)

While there is very little previous works to rely on, we do expect that at least some values will be connected to some 
%concrete aspects of songs (that we will call "attributes" in this paper). 
music attributes.
In order to assess which links would be relevant for future research and applications, we need to answer the two following questions for each potential link.

\begin{itemize}    

\item Research Question 1.1: How strong is the link?

Links between values and musical attributes or consequences need to be strong enough to be useful. Indeed, if Value "Ambition" is only very slightly supported by the Attribute "Loud", this may not be the most interesting link to investigate further, nor the most interesting in an application perspective. The strength, or intensity, of the links will be measured by the average score given by the participants. The further from 0, the stronger the link will be. 
This average score will also inform about the directions of the link : whether it promote or demotes the value.

\item Research Question 1.2: How stable is the link? 

Links between values and musical attributes or consequences also need to be stable across the population. In other words, most people have to agree about the link's existence and strength. Concretely, we will measure stability in this experiment by looking at the standard deviation of the ratings for each possible link. The smaller it will be, the more stable the link will be.
\end{itemize}

We do not have any hard hypothesis as to what those links will be because of the lack of previous research on the topic. This work's main objective is to provide a first look at which links are more or less stable across participants. Thus also providing directions for deeper research on this topic by identifying the links that are worth investigating further.

\item Research Question 2: Can we directly link personal values and songs attributes without going through the Consequences step ?
%Is the middle-step consequences level needed to build mappings between values and musical preferences and to infer those preferences from someone's values? (see marketing paper)

According to \cite{van1994domain}, abstract personal values and concrete product attributes usually need to be connected by an intermediary abstraction level. However, the study that originally made this claim could not be found. As a result, we want to verify this claim in our context. Using this extra level makes the procedure more tedious for participants and less straightforward. In the perspective of future research and applications of our method, knowing whether this extra step is indeed necessary will be a useful contribution.

We will answer this question by comparing the outcomes of two experimental conditions: a direct one without this intermediary step and an indirect one with it.

\end{itemize}

\subsection{Potential Applications}
%Either before or after the RQ

%\subsection{Recommender Systems}
A better understanding of the links between personal values and musical attributes would have interesting applications in the recommender systems field. Recommender systems are algorithms designed to help users navigate a vast, and often ever-growing, collection of items. Traditionally, they rely on users past history on the system. What did they consume? What did they like? The two main approaches are collaborative filtering and content-based. Collaborative-filtering methods use the previous interactions of users and items to compute the similarity between users or between items. It then recommends something that users with a similar consumption to the current user liked; or an item similar to the ones the current user previously liked. Content-based methods use explicit information about what the items are. For example, for a song, it would be artist, year of release, genre, country, ... It then recommends items with similar characteristics than the ones the user liked or consumed before \cite{Ricci2015}. 

Therefore, music recommender systems often focus on what its users previously listened to, mostly directing them towards similar songs. Part of the field have focused on improving recommender systems by directing them towards more surprising, but still good, recommendations. We believe that using personal values in recommender systems will help to provide taste expanding recommendations that are still tailored for each particular user. Indeed, such recommender will not rely (or at least less) on what users have already listened to. On top of this, personal values representing who people want to be, users might be intrinsically motivated to follow personal values based recommendations. 

Our line of research aims to explore this by looking into the strength and stability of the links between personal values and musical attributes. Our results will provide insights about the potential usefulness of personal values in the context of music recommendations.

\section{Methods}

This study is composed of several steps. First, we conducted a short pilot study to test the survey. Then, we conducted a large simulation study to both prepare our analysis and estimate the number of participants we would need. Finally, we plan to pass the survey to a sample of the general population via Prolific, a British crowdsourcing platform designed for scientific research.

\subsection{Pilot}

%The survey went through several rounds of refinement based on feedback received from pilot participants. Those were mainly recruited among our colleagues. Our main questions were the understandability of the survey as well as its length and cognitive load. Once we were satisfied with the feedback, the survey went to one last round of pilot to gather 5 results per conditions in their final versions. This number of results is not enough to draw any conclusion but serves as a basis for the simulation.

Before running the simulation and the experiment, we conducted a pilot study. Our goal was to pre-test the survey, making sure the instructions were understandable and that it was not too long. For this, we added a couple of questions at the end of the survey, as well as an open question for eventual feedback. We also wanted to know how long on average people took to complete it in order to calculate the appropriate reward for the participants. Our second goal was to gather some estimate of the parameters required to run our simulation.

Indeed, the simulation requires an estimate of the population parameters in order to run properly. Those parameters can be estimated from previous similar studies or through a pilot study. As we could not find relevant previous studies on our topic, we had to use a pilot.
From this pilot, we only extracted the information needed for the simulation: standard deviation of the ratings for each link, standard deviation of the ratings of each participant and the mean of the differences in the stability of the ratings (standard deviation) between the two conditions. 

The pilot was run on 2 of our colleagues who were not a part of this project. We then recruited 8 participants via Prolific to achieve a sample of 10 participants. Because of the random attribution to one of the two conditions, 6 went through the indirect condition and 4 through the direct condition. Each condition had a different median completion time : respectively 51:05 minutes and 16:45 minutes. 
According to this and due to Prolific reward system, it appears that the best way to run our study would be to run a pre-screening study to select participants who don't mind going through either condition. Then, use their Prolific IDs to randomly assign them to either condition, that would be run as 2 separate studies on the Prolific platform in order to pay participants fairly or their time. We will reward participants based on the median completion time of the conditions they will be assigned to and a £9 per hour rate \footnote{This is the fair hourly rate recommended by Prolific.}. 

The analysis \footnote{see "Pilot Analysis.Rmd" file in the code joined with this paper.} showed an average standard deviation of the ratings per link of 0.41 and a standard deviation of the ratings per participant of 0.49. 

%An Intraclass Correlation Coefficient was run for each condition to assess raters agreement. The results showed a moderate agreement in both conditions (0.67 for the direct condition and 0.712 for the indirect condition)  \cite{koo2016guideline}, using the two-way random effect models and the average of the ratings, p $<$ .001.

\subsection{Simulation Study}

After and along with the pilot, we built a simulation study. Simulation studies have several advantages. First, they allow writing the script of the analysis in advance. This allows to spot ahead of time changes to be made 
%and errors to correct 
before passing the survey. A second advantage is that it allows to get an estimate of the number of participants needed by running the simulation several times and comparing the outcomes with the pre-established effect and population parameters. A third advantage is that it also allows testing the planned analysis and spot the measures and statistical tests that may not be appropriate in practice. This is especially relevant where no well-established, straightforward method already exists to answer the research question based on the data collected, as it is the case here.

The simulation's code was written in R and based on a tutorial \cite{debruine2021understanding}.\footnote{The detailed and commented code notebooks can be found at: \url{https://github.com/smanolios/validation-values-music-map} and has been joined with this paper as well.} In summary, the code return a simulated set of ratings by simulated participants and takes as input population and effect parameters: 
\begin{itemize}
    \item n\_subj: the number of total participants to simulate
    \item $\beta_0$: the grand mean of all the score from the "ground truth". 
    \item $\tau_0$: standard deviation of the by-subjects random intercept. This simulates the differences in rating behavior among participants. For example, some use all the length of the scale, while some almost never use the more extreme values.
    \item $\omega_0$: standard deviation of the by-items random intercept.This simulates the differences between the links. Some may be more obvious than others for most participants, and would then have a smaller standard deviation than other links. This parameter account for this effect.  
    \item $X$: the raw average difference in rating between the conditions. This parameter simulates any difference in the ratings due to the effect of conditions. For example, if raters give more extreme ratings in the indirect condition than in the direct condition.
    \item $\sigma$: standard deviation of the residual error. Because people are complex and not everything can be controlled, there will always be an inexplicable small amount of variation left. This is what this parameter simulates. 
    
\end{itemize}

Based on those parameters, the script simulates the items (links) as well as the desired number of participants. It assigns the first half of the participants to the first condition (direct) and the second half to the second condition (indirect). The ratings each simulated participant gives to each item (link) in their experimental condition is then calculated with the following equation: 

\begin{math}score = \beta_0 + \tau_{0s} + O_{0i} + X_i + e_{si}\end{math}

where $O_{0i}$ is the random intercept for item \emph{i}, $X_i$ the condition effect on item \emph{i} and $e_{si}$ the residual error of the trial involving subject \emph{s} and item \emph{i}.

In the absence of similar past works to rely on, we assume by default that the ratings will follow a normal distribution. We will check if this assumption is correct once we have gathered enough data.%We based this assumption on the results of the pilot study.
%Add citation

\subsection{Simulation Analysis}

We conduct different analysis of the simulation results. We first computed the Cohen $d$ effect size for each simulation to get a standardized and therefore interpretable and comparable effect size.  

The second one is the same as in the tutorial : conduct a power analysis to see how the power varies based on the effect of the condition and the error. For this, we simply run a linear model using the following equation :

$score = condition+(1|participant_id)+(1|item_id)$

and extract the p-values. Then, for each bulk of 1100 simulations, we simply count which proportion of p-values is under the alpha threshold. For more flexibility, we use both 0.05 and 0.01 as alphas. 

We also performed the same procedure on scaled scores. In the real experiments, participants will only be able to give ratings between -1 and 1, but the simulation does not have the same boundaries when generating the scores. We corrected this using a sigmoid function. We kept both scaled and unscaled scores, as both have their pros and cons.

The third analysis compare the raw distributions of the mean of each link's ratings between the two conditions using two single sided t-tests complemented by an equivalence test (with a minimal effect size of 0.2). We test for both difference and equivalence to distinguish if an effect is significant, too small to be of interest / non-existent or in a non-interpretable gray area.  
We picked the minimal effect size of interest of 0.2 as it is the suggested threshold between a negligible and a small effect according to Cohen $d$'s interpretation guideline \cite{cohen1992quantitative}.

\subsection{Simulation Iteration Process}

We ran 1100 simulations using the following parameters : 
\begin{itemize}
    \item $\beta_0$ = 0
    \item $\tau_0$ = 0.486
    \item $\omega_0$ = 0.410
    \item number of subjects = 100, 200, 400, 600, 800, 1000
    \item condition effect = 0, 0.024,0.05,0.1, 0.5, 0.8
    \item $\sigma$ = 0.1, 0.2, 0.5, 0.8 and 1
\end{itemize}

 The parameters $\tau_0$ and $\omega_0$ (and condition effect for the 0.024 value) are based on the pilot results and $\sigma$ is a degree of freedom for which we will test several values.

We used a linear regression model on the outcome of each simulation in order to collect the p-values needed to compute the overall power of each batch of 1100 simulations. The following equation was used : 

$link\_rating = \beta_0 + (1|participants) + (1|items)$ 

where both participants and items are considered as random effects. Indeed, we only have a sample of the population and each participant differ from the next in many ways, including their usage of the scale from the survey. And the list of elements (Attributes and Consequences) and therefore of the pairs used in this experiment is not necessarily exhaustive. Plus, some pairs may be more obvious to rate than others.

\subsection{Simulations Results}

\begin{figure}[h]
\caption{\textbf{Morey Plots of the results of the simulations}. Those plots show the evolution of power depending on number of participants, error and effect of condition. Each plot represent a different effect size. This time, the effect of condition is not raw but translated into a Cohen $d$. The horizontal line represents the 90\% power and the vertical lines the conventional cut-offs between the different interpretations of the effect size (negligible, small, moderate, large).}
\centering
\includegraphics[scale=0.70]{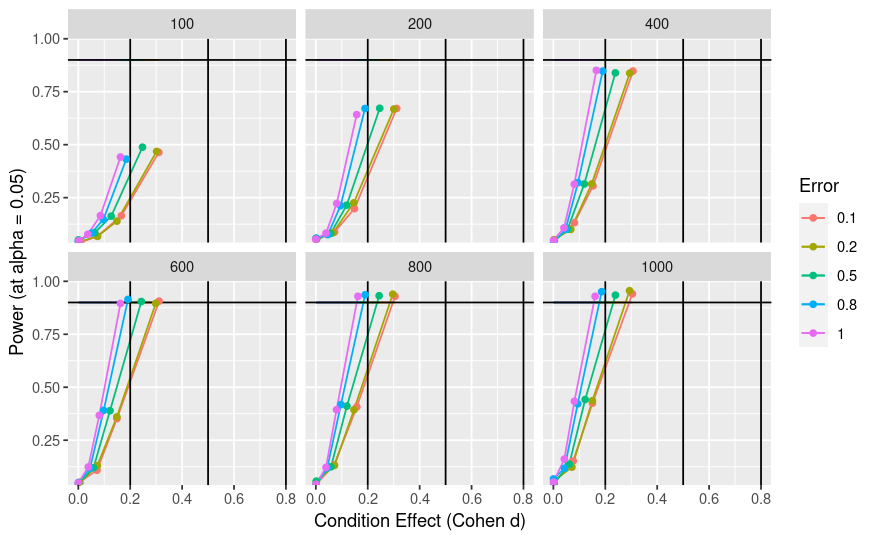} 
\label{fig:morey2}
\end{figure}

The main information we seek from those results is the estimation of how many participants we would need to achieve a reasonable amount of power. We choose to give our experiment 90\% power and decided that we are not interested in negligible effects. Therefore, we want our data points to be above the 90\% vertical line and on the right side of the 0.2 horizontal line. Based on this, we choose to get a sample of 600 participants.

\subsection{Population}

We will recruit our participants via Prolific. Based on the simulation result, we will recruit our 600 participants with only two criteria: they should be 18 years old or older (for ethical reasons) and be fluent in English (to be able to understand the survey properly). They will be compensated upon completion of the survey on the basis of 9 pounds an hour \footnote{This is equivalent to 10.32 euros or 10.4 US dollars}. Based on the estimated time of completion of each condition from the pilot, our study will therefore cost around £3050 \footnote{This is equivalent to 3470.44 euros or 3588.48 US dollars}.
%300×(9÷60×51)+300×(9÷60×16,75)

\subsection{Ethics}

This experiment has already been approved by Delft University of Technology's Ethics Committee. Our Data Management Plan has received approval as well. In short, participants will be informed of how their personal information and answers will be treated. They will have to agree before participating in the experiment. Demographic information will be reported in an aggregated level and will not be shared outside the research team to respect our participant's privacy. Their ratings of the links will be made available online in a dataset that will be provided with the complete version of this work. Only an anonymized participant ID will be connected to the ratings. 

\section{Materials}

\subsection{Survey}

The survey was built based on the results of a previous interview study \cite{manolios2019influence}. It used the Attributes and Consequences elicited during those interviews, minus those that were deemed not relevant in this context (because not informative enough), such as "Genre" or "Features of Music" \footnote{The complete lists of keywords for this study is available in the supplementary materials.}. 

\begin{figure}[h]
\caption{Survey Flow}
\centering
\includegraphics[scale=0.25]{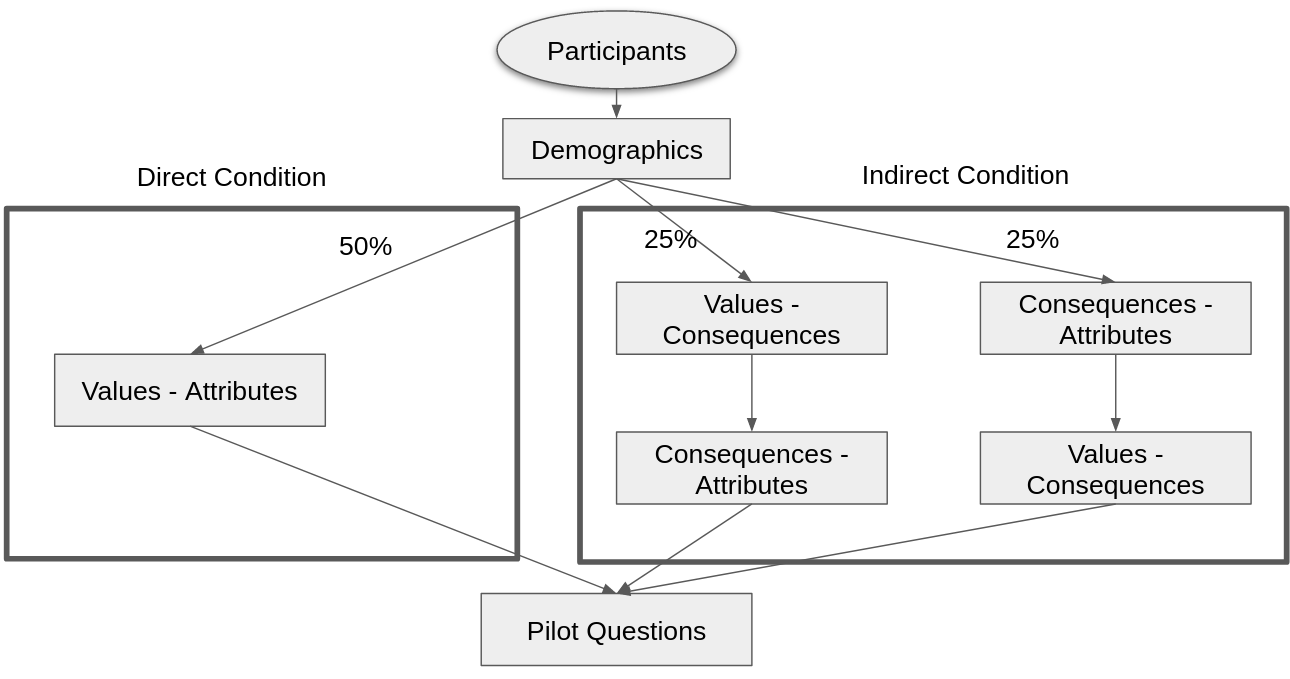}
\label{fig:surveyflow}
\end{figure}

The survey is composed of a first standard demographics' section in which we collect participants' ages, declared gender and country in which they grew-up. Those questions' only purpose is to assess our sample's diversity. We ask for the country in which participants grew-up specifically, instead of their nationality or country of origin, because what kind of music people are exposed to during their teenage years weight a lot in the formation of their musical preferences (see Section \ref{musicpsycho}).  We also ask about which languages participants are fluent in to verify that English is one of them.

\begin{figure}[h]
\caption{Scale example from the survey}
\centering
\includegraphics[scale=0.25]{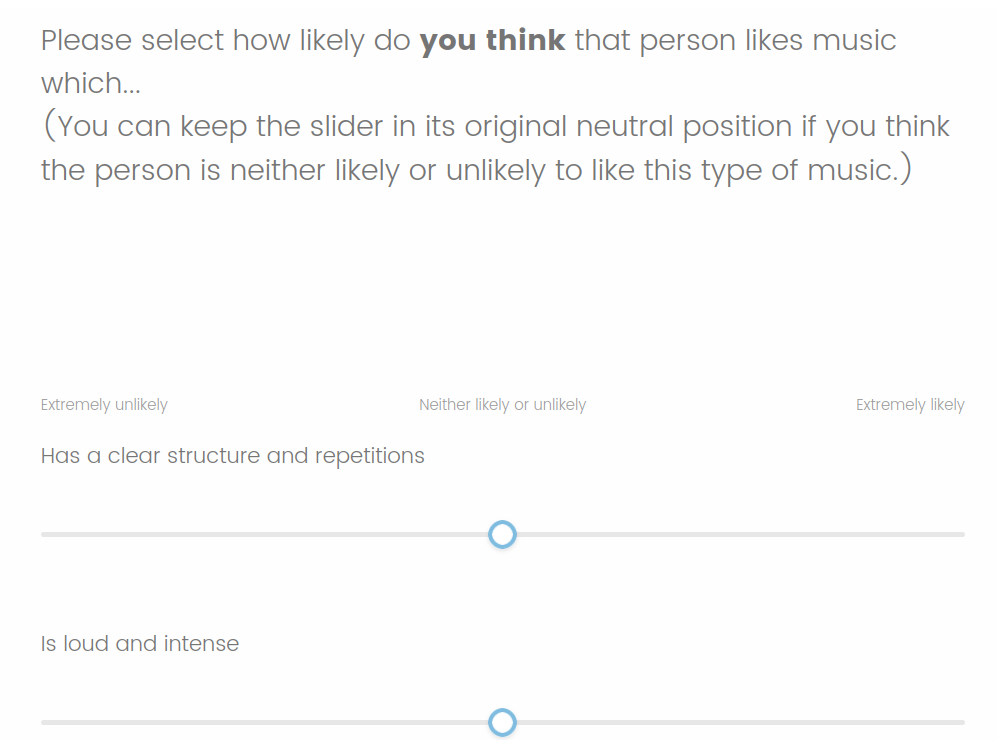}
\label{fig:surveyscale}
\end{figure}

Then, the survey breaks down in two conditions (to answer RQ2), as shown in Figure \ref{fig:surveyflow}. In the first condition, participants have to directly link personal Values to musical Attributes. Values are represented by a value portrait: a description of a hypothetical person's values. Those portraits were built by stacking the items of Schwartz's Portrait Value Survey (PVQ) \cite{Schwartz2005, Schwartz2001}, with a minimal amount of modifications designed to unify the sentences and smooth the text. No keyword were modified \footnote{The complete text for each portrait is available in the supplementary materials.}. For each of those portraits, the following question is asked: "How likely do you think this person is to enjoy music that is..." Under this question a list of continuous scales with no visible numeric values are presented, ranging from "Extremely Unlikely" to "Extremely likely" with a "Neither likely nor unlikely" as the middle and starting point as shown in Figure \ref{fig:surveyscale}. Each scale is associated with each of the 10 Attributes. 
We choose to use continuous scales to allow for more fine-grained data collection and analysis \cite{toepoel2018sliders,chyung2018evidence}. We used a neutral middle point because we don't expect that each Attribute is relevant for each Value. This middle point reduces the cognitive load of participants by allowing them to simply ignore the scale / question if they don't think this link exist. They then have less scales to rate, and at the same time we get a clear signal that the participant doesn't think this particular Attribute - Value association is relevant.

The second condition contained two sets of questions, presented in random order, as shown on the right part of Figure \ref{fig:surveyflow}. One set have participants connect Values to Consequences, in the same way that participants of the first condition connect Values to Attribute. The other set of question is designed to have participants connect Consequences to Attributes. Participants are asked: "To which extent do you agree with the following statements? Music that is [insert Consequence] is also:". This question is followed by a series of continuous scales without numerical values labeled "Extremely Disagree" to "Extremely Agree" with "Neither Agree nor Disagree / Not Applicable" as the middle and starting point. The scale format and the rationale behind it are the same as the ones of the other questions. Only the labels has been slightly modified to remain grammatically coherent with the question phrasing. Using the same scale will be useful for the analysis, especially to be able to draw comparisons between the questions. This advantage was important enough to balance the disadvantage of making the second condition longer and more fastidious than a simple binary answer format, which was our original idea.

In both conditions, the order of the Values, Attributes and Consequences was randomized, and each participant was presented with each possible Value-Attribute pair in the first condition or each possible Value-Consequences and Consequences-Attribute pair in the second condition. This randomization allows controlling any unwanted effect of the order in which the items are presented.

\section{Analysis Plan}

The code for the analysis has been written in R Markdown and is entitled Analysis.Rmd. It aims to be as readily understandable as possible. The functions are in a separate Functions.Rmd file.\footnote{The code is joined to this paper.}

\subsection{Preprocessing}

We will first exclude from the experiment results from incomplete surveys and from participants who did not include English in any of the language questions from the Demographics section. 

Across all surveys, we expect a certain number of links to be considered nonexistent or non-relevant by the participants. We asked in the instructions of each question to leave the cursor at its starting position on the scale in this case. But Qualtrics considers those cases as NAs, while in our context, they are actually 0s ("Neither likely nor Unlikely"). Therefore, all NAs ratings in each participant's assigned condition will be changed to 0.

\subsection{Demographics}

We will only use the demographic questions as a way to assess our sample's diversity. Therefore, we will only compute and report the mean and standard deviation of our participant's ages and the division of our participant across gender and country in which they grew up. We will not look for any potential effect of those variables on the survey answers, as this is quite far from our research questions and would call upon a different study and research design.

\subsection{RQ1: What are the links between personal values and musical preferences?}

In order to answer this question, we will look at participants ratings of each of those links in two ways: $\mathbf{intensity\ of\ the\ effects}$ and $\mathbf{raters\ agreement}$. 

\subsubsection{RQ1.1 : How strong are the links?}
The former will give insight on how strong, and therefore interesting, the effect is. For this, we will compute the mean of the ratings of each link. The further it will be from 0, the stronger the effect would be. Closer to 1 it will be a positive effect: for example, the Value A is promoted by Musical Attribute B. If the mean rating would be closer to -1 then the Value A would be demoted by the Musical Attribute B. 

\subsubsection{RQ1.2 : How stable are the links?}
The latter will give insight on how stable the effect is. Do most participants tend to agree on the presence of a link, its direction and its intensity? Many measures of raters agreement exist, with little agreement on which one should be used in a specific context. Most of them are across items, but to answer our research question, we need a measure of agreement that can be computed for each link (item) individually. Therefore, we decided to use the standard deviation of each link's ratings. We consider it the most appropriate one, as it reflects a very clear picture of how spread out the ratings are from the mean.

In a slightly exploratory perspective, we will also compute the ICC (Interclass Correlation Coefficient) to have a more global picture. ICC is a metric that computes global rater agreement across all the items. It will give some insight whether, in general, people tend to agree on the links between musical preferences and personal values. 

Due to a lack of previous studies to rely on, there is no clear signal on which cut-off should be used to decide if a link is strong enough to be considered interesting and if the agreement is high enough to consider this representation of the link is stable among our sample. Therefore, we will rank the links and interpret the outcomes of this analysis in a more qualitative way to shed some light on which links are the strongest and most stable compared to the others. Those will be the most promising links to investigate further in later studies.

%\subsection{RQ2: Comparison of both interviews' and survey's maps}

%In this step, we will use the musical preferences and values map from \cite{manoliosinfluence2019}. In this qualitative study, the questions were about the participants' own declared musical preferences and how they connect to their values. Therefore, the map represents the participants' self-representations of the links we investigate. This current work however asks about hypothetical other persons and is then about participants' more global representations of how musical attributes link to personal values. It is then interesting to compare both and get some insight on whether those self-representations and global representations are similar or not. 

%Because the self-representations map was obtained by qualitative means, this step of the analysis will also be conducted in a qualitative way. We will use a matrix of the intensity and stability of the links from the previous step. Then, we will compare it to the one from the Supplementary Materials of \cite{manoliosinfluence2019} to look at which links are present in both, or at the contrary present in one but not the other.

\subsection{RQ2: Can we directly link personal values and song attributes?}

The two conditions of this experiment are designed to answer this research question. We want to know whether, as claimed in \cite{van1994domain}, an intermediate abstraction step is needed to connect concrete Attributes and Values, in our musical preferences' context. 
In order to assess this, 
%we will first normalize the ratings using the z-scores of the participants to account for the participant effect. Then, we will compare the distributions of the means of the ratings in each condition using a combination of t-test and equivalence test on the absolute values. 
%we will  compute the Interclass Correlation Coefficient for each condition and look at whether the 95\% confidence interval overlap to see whether there is a significant difference in raters agreement. We will use a two-way model computing the consistency (versus absolute agreement) of the raters, which takes into account the possibility that raters use the scale differently. This test will give a first insight on potential differences in difficulty of the tasks. In other words, if it is harder to connect elements when they are separated by more than one level of abstraction.
we will compare participants absolute ratings in both conditions using two one-sided t-tests on the absolute scores, and an equivalence test. Those tests will give a first insight in potential differences in absolute rating distribution between the two conditions. The equivalence test will serve to measure whether the effect is too small to be of interest.

%In a second step of analysis, %we will repeat the steps from RQ1 to identify the most stable and strongest links. We will then qualitatively compare the paths between Values and Attributes between both conditions. The goal there will be to see if we can find the same connections between Values and Attributes

% Results and Discussion can be combined.

%\section*{Conclusion}

\section*{Acknowledgments}

We would like to thank everyone who helped us throughout this project. In particular, Prof. Dr. Nava Tintarev for her precious input at the beginning of this project and Dr. Chelsea Parlett Pelleriti for her very valuable insights on the statistical parts of this work.

\section*{Supplementary Materials}
\subsection{Item List}
\subsubsection{Attributes}
\begin{itemize}
    \item Is fun
    \item Gives Energy
    \item Allows them to express their identity
    \item Provokes negative feelings
    \item Support creativity
    \item Is familiar and evokes old memories and nostalgia
    \item Is unique
    \item Makes them feel powerful
    \item Makes them feel something
    \item Is challenging and not boring
    \item Makes the discover something new
    \item Has meaning and authenticity
    \item Connects people
    \item Help them focus and be more productive
    \item Shows the efforts, talents and skills of the artists
    \item Help them improve themselves
    \item Is relaxing
\end{itemize}

\subsubsection{Consequences}
\begin{itemize}
    \item Expresses an emotion
    \item Is original
    \item Is harmonious and melodic
    \item Is simple and accessible
    \item Has a story
    \item Has a clear structure and repetitions
    \item Is loud and intense
    \item Is improvised and spontaneous
    \item Has a soft sound
\end{itemize}

\subsubsection{Portraits}
\begin{itemize}
    \item Conformism : "They want to avoid doing anything people would say is wrong, and they believe that people should do as they are told and follow rules at all times, even when no one is watching. They try to never disturb or irritate others. It is important to them to behave properly and to be obedient and polite to other people all the time, as well as show respects to their parents and older people."
    \item Tradition : "It is important to them to keep the customs they have learned, and they think it is best to do things in traditional ways. Religious belief are important to them, and they try hard to do what their religion requires. They think it's important not to ask for more than they have and that people should be satisfied by it. It is also important to them to be humble and modest, and they try not to bring attention to themselves."
    \item Benevolence : "They want to devote themselves to people close to them : they deem very important to help the people around them, and they want to care for their well-being. Loyalty to their friends and respond to the needs of others is also important to them, and they try to support those they know. It is also important for them to forgive people who have hurt them. They try to see what is good in others and not to hold a grudge."
    \item Universalism : "They believe all the world's people should live in harmony and they deem important to promote peace among all groups in the world. They want everyone to be treated justly, even people they don't know. They think it is important that every person in the world be treated equally and that the weak in society are protected. They believe everyone should have equal opportunities in life. They seek to understand people, even those they disagree with. They think it is important to listen to people who are different from themselves. Looking after the environment is important to them, as well as to adapt to nature and fit to it. They strongly believe that people should not change nature and care for it."
    \item Self-Direction : "Being independent is important to them. They like to do things in their own original ways. It is important to them to think up new ideas and to be creative, as well as making their own decisions about what they do. They like to rely on themselves and to be free to plan and to choose their activities for themselves. They also like to be curious and to try to understand all sorts of things and they think it is important to be interested in things. "
    \item Stimulation : "It is important to them to have an exciting life. They like surprises and always look for new things to try. They are always looking for adventures, and they like to take risks. They think it is important to do lots of different things in life. "
    \item Hedonism : "They really want to enjoy life. Having a good time and enjoying life's pleasure is important to them. They like to 'spoil' themselves and seek every chance they can to have fun. It is important to them to do things that give them pleasure."
    \item Achievement : "It is important to them to show their abilities and they think it is important to be ambitious. Being very successful  and getting ahead in life are important for them. They strive to do better than others and they like to impress people. They want to show how capable they are and people to admire what they do. "
    \item Power : "They like to be the leader and always want to be the one who makes the decisions. Being in charge and tell others what to do is important for them : they want people to do what they say. It is also important for them to be rich : they want to have a lot of money and expensive things."
    \item Security : "Living in a safe country and secure surroundings is important to them. They avoid anything that might endanger their safety and think the state must be on watch against threats from within and without. Having a stable government is important to them. They are concerned that the social order be protected. It is important to them that things are organized and clean. They really don't like things to be a mess. They also try hard to avoid getting sick, as staying healthy is very important to them as well."
\end{itemize}

%\nolinenumbers

\bibliography{biblio.bib}
% Either type in your references using
% \begin{thebibliography}{}
% Text
% \end{thebibliography}
%
% or
%
% Compile your BiBTeX database using our plos2015.bst
% style file and paste the contents of your .bbl file
% here. See http://journals.plos.org/plosone/s/latex for 
%\begin{thebibliography}{10}
%
%\bibitem{bib1}
%Conant GC, Wolfe KH.
%\newblock {{T}urning a hobby into a job: how duplicated genes find new
%  functions}.
%\newblock Nat Rev Genet. 2008 Dec;9(12):938--950.
%
%\bibitem{bib2}
%Ohno S.
%\newblock Evolution by gene duplication.
%\newblock London: George Alien \& Unwin Ltd. Berlin, Heidelberg and New %York:
%  Springer-Verlag.; 1970.
%
%\bibitem{bib3}
%Magwire MM, Bayer F, Webster CL, Cao C, Jiggins FM.
%\newblock {{S}uccessive increases in the resistance of {D}rosophila to %viral
%  infection through a transposon insertion followed by a {D}uplication}.
%\newblock PLoS Genet. 2011 Oct;7(10):e1002337.
%
%\end{thebibliography}

\end{document}